\documentclass[twocolumn,aps,pra,amsmath,amssymb]{revtex4-1}
\usepackage[latin9]{inputenc}
\usepackage{amsmath}
\usepackage{amssymb}
\usepackage{graphicx}
\usepackage{esint}
\usepackage[unicode=true,
 bookmarks=false,
 breaklinks=false,pdfborder={0 0 1},backref=false,colorlinks=false]
 {hyperref}
\usepackage{breakurl}

\makeatletter

\newcommand*\LyXThinSpace{\,\hspace{0pt}}

\usepackage{epsfig}
\usepackage{bm}
\usepackage{dcolumn}
\usepackage{color}

\makeatother

\begin{document}

\title{Role of the confinement-induced effective range on the thermodynamics
of a strongly correlated Fermi gas in two dimensions}

\author{Brendan C. Mulkerin$^{1}$, Hui Hu$^{1}$ and Xia-Ji Liu$^{1}$ }

\affiliation{$^{1}$Centre for Quantum and Optical Science, Swinburne University
of Technology, Melbourne, Victoria 3122, Australia}

\date{\today}
\begin{abstract}
We theoretically investigate the thermodynamic properties of a strongly
correlated two-dimensional Fermi gas with a confinement-induced negative
effective range of interactions, which is described by a two-channel
model Hamiltonian. By extending the many-body $T$-matrix approach
by Nozi\`eres and Schmitt-Rink to the two-channel model, we calculate
the equation of state in the normal phase and present several thermodynamic
quantities as functions of temperature, interaction strength, and effective
range. We find that there is a non-trivial dependence of thermodynamics
on the effective range. In experiment, 
where the effective range is set by the tight axial confinement, the
contribution of the effective range becomes non-negligible as the temperature
decreases down to the degenerate temperature. We compare our finite-range
results with recent measurements on the density equation of state,
and show that the effective range has to be taken into account for
the purpose of a quantitative understanding of the experimental data.
\end{abstract}

\maketitle

\section{Introduction}

Recent developments in creating ultracold atomic gases in lower dimensions
has opened up a new and exciting field to explore strongly correlated
many-body systems \cite{Levinsen2015,Turlapov2017}. Two-dimensional
(2D) ultracold atomic Fermi gases are of interest due to the increased
role of thermal and quantum fluctuations, which lead to, for example,
suppressed superfluid long-range order at nonzero temperature and
a quasi-ordered transition to superfluidity, the Berezinksii-Kosterlitz-Thouless
(BKT) transition \cite{Berezinski1972,Kosterlitz1973}. In addition
to presenting a range of novel and intriguing quantum phenomena, 
such as scale invariance and the breathing mode anomaly
\cite{Holstein1993,Pitaevskii1997,Olshanii2010,Hofmann2012,Taylor2012}
and novel topological phases \cite{Liu2012,Cao2014,Goldman2016},
2D Fermi gases provide an important tool in understanding confined
many-body systems from diverse fields of physics, such as: high-temperature
superconductors \cite{Loktev2001}, thin $^{3}$He films \cite{Ruggeri2013},
neutron stars in a nuclear pasta phase \cite{Pons2013}, and exciton-polariton
condensates in a microcavity \cite{Deng2010}.

The key advantage of 2D ultracold Fermi gases is their unprecedented
controllability: the interatomic interaction can be tuned continuously
with Feshbach resonances to realize the crossover from a Bardeen-Cooper-Scrieffer
(BCS) superfluid of weakly interacting Cooper pairs to a Bose-Einstein
condensate (BEC) of tightly bound molecules \cite{Turlapov2017,Makhalov2014},
the population of spin components can be changed \cite{Feld2011},
the confining potential can be made homogeneous with a box potential
\cite{Hueck2018}, and non-abelian synthetic gauge field (i.e. spin-orbit
coupling) can be devised \cite{Zhang2014}.

With these advancements, a set of seminal experiments directly probed
the universal thermodynamics of a 2D interacting Fermi gas at given
interaction strengths \cite{Makhalov2014,Martiyanov2016,Fenech2016,Boettcher2016},
by measuring the in-trap density profile. This allows for a direct
comparison between theoretical predictions and experimental data.
To date, it has been found that, a single-channel model with a single
2D $s$-wave scattering length works reasonably well in describing
the equation of state (EoS) experimental results, both in the normal
state \cite{Watanabe2013,Bauer2014,Barth2014,Anderson2015,Mulkerin2015}
and below the BKT transition \cite{Bertaina2012,He2015,Shi2015,Mulkerin2017}.
However, in the strongly correlated regime it turns out that, the
theoretical EoS determined by accurate auxiliary-field quantum Monte
Carlo (AFQMC) simulations \cite{Shi2015} slightly over-estimates
the thermodynamics compared to the measured EoS \cite{Turlapov2017,Makhalov2014,Martiyanov2016},
suggesting the inefficiency of the single-channel model. The necessity
of using a more appropriate theoretical model was highlighted by the
most recent measurements on the breathing mode \cite{Vogt2012,Holten2018,Peppler2018},
which present an interesting example of a quantum anomaly (i.e., violation
of the classical scale invariance \cite{Pitaevskii1997}). It was
found that the measured breathing mode is \emph{notably} lower than
the prediction from the single-channel model \cite{Hofmann2012,Taylor2012}.
This discrepancy can not be fully understood by the nonzero but small
temperature found in the experiments \cite{Mulkerin2018}. Instead, it is
caused by a confinement-induced effective range of interactions, which
is negative and turns out be significant under the current experimental
conditions \cite{Hu2019}. By adopting a two-channel model to account
for the effective range, both measurements on low-temperature EoS
and breathing mode anomaly can now be satisfactorily explained by
 calculations at \emph{zero temperature} \cite{Wu2019}.

In this work, we explore how the confinement-induced
effective range changes the \emph{finite-temperature} thermodynamic
properties of a normal, strongly-correlated 2D Fermi gas. Our purpose
is two-fold. First, characterizing the thermodynamics of the normal
state of strongly correlated Fermi gases may be relevant to understanding
the role of many-body pairing \cite{Murthy2018}, which is a precursor
to superfluidity \cite{Watanabe2013}. The results
presented here can be used to better understand the existing two experimental measurements
of the EoS at finite temperature in the normal phase \cite{Fenech2016,Boettcher2016}.
Through the simplest many-body $T$-matrix theory
developed by Nozières and Schmidt-Rink (NSR) \cite{Nozieres1985},
we determine the thermodynamic potential of the two-channel model,
and calculate how the pressure EoS, energy, and entropy change with
a \emph{negative} effective range. Second, we then consider the realistic
effective range in two recent experiments \cite{Fenech2016,Boettcher2016}
and compare our finite-range results with experimental data on density
EoS. We find that in the experiments the effect of the effective range
starts to show up when the temperature is decreased down to Fermi
degeneracy. To quantitatively explain the experimental density EoS
at finite temperature near superfluid transition, therefore, the effective
range has to be taken into account in future refined theoretical works.

We note that, for a three-dimensional (3D) interacting Fermi gas,
the effect of a negative effective range has been recently discussed
by Tajima \cite{Tajima2018,Tajima2019}, following the seminal work
by Ohashi and Griffin \cite{Ohashi2003}, who extended the NSR approach
to the two-channel model. There, the effective range is related to
the width of Feshbach resonance and the 3D interacting Fermi gas may
experience severe atom loss in the narrow resonance limit (i.e., at
a large negative effective range) \cite{Hazlett2012}. In our case,
the confinement-induced effective range is intrinsically set by the
tight axial confinement and the 2D interacting Fermi gas is always
mechanically stable.

Our manuscript is set out as follows. In Sec.~\ref{sec:hamil}, we
introduce the two-channel model Hamiltonian and renormalize the relevant
parameters by solving the two-body scattering problem. In Sec.~\ref{sec:Tmatrix},
we briefly outline the many-body effective field theory of the model
Hamiltonian and show how to calculate the thermodynamic observables.
In Sec.~\ref{sec:results}, we discuss the thermodynamic properties
of the 2D interacting Fermi gas, as functions of the effective range,
temperature and interaction strength. And finally in Sec.~\ref{sec:conc},
we summarize our results. For simplicity we set $\hbar=1$ throughout.

\section{Hamiltonian and two-particle scattering}

\label{sec:hamil}

We begin our calculation of the thermodynamic properties by considering
a two-component Fermi gas in the normal state, described by the two-channel
Hamiltonian \cite{Tajima2018,Tajima2019,Ohashi2003,Liu2005,Schonenberg2017}:
\begin{eqnarray}
\mathcal{H} & = & \sum_{\mathbf{k}\sigma}\xi_{\mathbf{k}}c_{\mathbf{k}\sigma}^{\dagger}c_{\mathbf{k}\sigma}^{\phantom{\dagger}}+\sum_{\mathbf{q}}\left(\epsilon_{\mathbf{q}}/2+\nu-2\mu\right)b_{\mathbf{q}}^{\dagger}b_{\mathbf{q}}^{\phantom{\dagger}}\nonumber \\
 &  & +g\sum_{kq}\left(b_{\mathbf{q}}^{\phantom{\dagger}}c_{\mathbf{q}/2+\mathbf{k}\uparrow}^{\dagger}c_{\mathbf{q}/2-\mathbf{k}\downarrow}^{\dagger}+{\rm H.c.}\right),
\end{eqnarray}
where ${\rm H.c.}$ is the Hermitian conjugate, $c_{\mathbf{k}\sigma}$
are the annihilation operators of fermionic atoms with spin $\sigma=\uparrow,\downarrow$
and mass $M$ in the open channel, and $b_{\mathbf{q}}$ the annihilation
operators of bosonic molecules in the closed channel. The kinetic
energy of atoms measured from the chemical potential $\mu$ is $\xi_{\mathbf{k}}=\epsilon_{\mathbf{k}}-\mu$,
where $\epsilon_{\mathbf{k}}=\mathbf{k}^{2}/(2M)$. The threshold
energy or detuning of the diatomic molecules is $\nu$ and the Feshbach
coupling between atoms and molecules is given by $g$.

In order to find the many-body properties we require the bare two-body
scattering parameters to be rewritten in terms of measurable or renormalizable
scattering parameters. To relate the detuning $\nu$ and the channel
coupling $g$ to physical observables, we consider the two-body $T$-matrix
in a vacuum ($E^{+}\equiv k^{2}/M+i0^{+}$) \cite{Hu2019,Liu2005},
\begin{equation}
T_{2B}^{-1}\left(E^{+}\right)=U_{\textrm{eff}}^{-1}+\sum_{\mathbf{p}}\frac{1}{2\epsilon_{\mathbf{p}}-E^{+}},
\end{equation}
where the effective 
interaction in the presence of the channel coupling is given by 
\begin{equation}
U_{\textrm{eff}}\left(E^{+}\right)=\frac{g^{2}}{E^{+}-\nu}.
\end{equation}
Taking a large momentum cut-off, $\Lambda\rightarrow\infty$, we write
the two-body $T$-matrix as 
\begin{equation}
T_{2B}^{-1}\left(E^{+}\right)=\frac{k^{2}/M-\nu}{g^{2}}+\frac{M}{4\pi}\left(\ln\left[\frac{\Lambda^{2}}{k^{2}}-1\right]+i\pi\right).
\end{equation}
Alternatively, we may rewrite it in the form,
\begin{equation}
T_{2B}\left(E^{+}\right)=\frac{m}{4\pi}\left(-2\ln\left[ka_{s}\right]-R_{s}k^{2}+i\pi\right),\label{eq:T2B}
\end{equation}
where we have written the detuning and Feshbach coupling in terms
of the 2D scattering length $a_{s}$ and effective range $R_{s}$
\cite{Schonenberg2017}, 
\begin{equation}
a_{s}=\frac{1}{\Lambda}e^{\frac{2\pi\nu}{gM}},\;\;R_{s}=-\frac{4\pi}{M^{2}}\frac{1}{g^{2}}.\label{eq:asRs}
\end{equation}
As can be seen from Eq.~\eqref{eq:asRs}, we recover the single-channel model
in the broad resonance limit when $g\rightarrow\infty$. We may remove
the cut-off $\Lambda$ by considering the pole of the two-body $T$-matrix
$T_{2B}(E)$, $E=E_{B}$. We find that, 
\begin{equation}
\nu=E_{B}+g^{2}\sum_{\mathbf{k}}\frac{1}{2\epsilon_{\mathbf{k}}-E_{B}}.
\end{equation}
The binding energy can be set by $\varepsilon_{B}=-E_{B}=\kappa^{2}/M$,
where $k=i\kappa$ corresponds to the pole of the two-body $T$-matrix. 

\section{The many-body $T$-matrix}

\label{sec:Tmatrix}

We consider the strong-coupling effects and pairing fluctuations by
applying the normal-state NSR approach \cite{Nozieres1985}, which
has been widely used in different context and has also been extended
to superfluid phase \cite{He2015,Hu2006,Hu2007,Diener2008}. For completeness,
here we briefly go through the derivation of the thermodynamic potential
using the functional path-integral formalism. All the thermodynamic
properties can be obtained through the thermodynamic potential, $\Omega=-k_{B}T\ln\mathcal{Z}$,
where the partition function $\mathcal{Z}$ for the two-channel Hamiltonian
is given by,
\begin{alignat}{1}
\mathcal{Z}=\int\mathcal{D}\left(c^{\,},c^{\dagger}\right)\mathcal{D}\left(b^{\,},b^{\dagger}\right)e^{-\mathcal{S}\left[c^{\,},c^{\dagger},b^{\,},b^{\dagger}\right]},
\end{alignat}
for atomic fields $(c_{\mathbf{k}\sigma}^{\phantom{\dagger}},\,c_{\mathbf{k}\sigma}^{\dagger})$
and molecular fields $(b_{\mathbf{q}}^{\phantom{\dagger}},\,b_{\mathbf{q}}^{\dagger})$,
and the action at the inverse temperature $\beta=1/(k_{B}T)$ is given
by,
\begin{alignat}{1}
\mathcal{S}=\int_{0}^{\beta}d\tau & \biggl[\sum_{\mathbf{k}\sigma}c_{\mathbf{k}\sigma}^{\dagger}(\tau)c_{\mathbf{k}\sigma}^{\,}(\tau)\nonumber \\
 & +\sum_{\mathbf{q}}b_{\mathbf{q}}^{\dagger}(\tau)\partial_{\tau}b_{\mathbf{q}}^{\,}(\tau)+\mathcal{H}(\tau)\biggl].
\end{alignat}
Using the Hubbard-Stratonovich transformation to decouple the Feshbach
coupling term and integrating out the atomic fields, we obtain an
effective action for Cooper pairs and molecules. By further truncating
the perturbative expansion over bosonic fields (for both pairs and
molecules) at the Gaussian fluctuation level, we arrive at the NSR
thermodynamic potential \cite{Ohashi2003},
\begin{equation}
\Omega=\Omega_{{\rm F}}+\Omega_{{\rm B}}-\sum_{\mathbf{q},i\nu_{n}}\ln\left[1+g^{2}D_{0}(\mathbf{q},i\nu_{n})\Pi(\mathbf{q},i\nu_{n})\right],
\end{equation}
where $\nu_{n}=2n\pi/\beta$ are the bosonic Matsubara frequencies,
$\Omega_{{\rm F}}=2\sum_{\mathbf{k}}\ln(e^{-\beta\xi_{\mathbf{k}}}+1)$
and $\Omega_{{\rm B}}=\sum_{\mathbf{q}}\ln(e^{-\beta\epsilon_{\mathbf{q}}^{{\rm B}}}-1)$
are the free fermionic and bosonic thermodynamic potentials for atoms
and molecules, respectively. We have defined the pair correlation
function 
\begin{alignat}{1}
\Pi(\mathbf{q},i\nu_{n})=\sum_{\mathbf{k}}\frac{1-f(\xi_{\frac{\mathbf{q}}{2}-\mathbf{k}})-f(\xi_{\frac{\mathbf{q}}{2}+\mathbf{k}})}{2\epsilon_{\mathbf{k}}-2\mu+\epsilon_{\mathbf{q}}/2-i\nu_{n}},
\end{alignat}
and $D_{0}(\mathbf{q},i\nu_{m})=1/(i\nu_{n}-\epsilon_{\mathbf{q}}^{{\rm B}})$
is the Green's function of a free molecular boson with dispersion
$\epsilon_{\mathbf{q}}^{{\rm B}}=\mathbf{q}/2-\nu+2\mu$. The thermodynamic
potential can be rewritten into the following form: 
\begin{equation}
\Omega=\Omega_{{\rm F}}-\sum_{\mathbf{q},i\nu_{n}}\ln\left[-\Gamma^{-1}(\mathbf{q},i\nu_{n})\right],
\end{equation}
where we have introduced the vertex function 
\begin{equation}
\Gamma^{-1}(\mathbf{q},i\nu_{n})=U_{{\rm eff}}^{-1}\left(\mathbf{q},i\nu_{m}\right)+\Pi\left(\mathbf{q},i\nu_{m}\right),
\end{equation}
and the in-medium effective interaction $U_{{\rm eff}}(\mathbf{q},i\nu_{m})\equiv g^{2}D_{0}(\mathbf{q},i\nu_{m})$,
which can be explicitly written as,
\begin{eqnarray}
\frac{1}{U_{{\rm eff}}\left(\mathbf{q},i\nu_{m}\right)} & = & -\sum_{\mathbf{k}}\frac{1}{2\epsilon_{\mathbf{k}}+\varepsilon_{B}}-\nonumber \\
 &  & \frac{M^{2}R_{s}}{4\pi}\left(i\nu_{n}-\frac{\epsilon_{\mathbf{q}}}{2}+2\mu+\varepsilon_{B}\right).\label{eq:Ueff}
\end{eqnarray}
Using Eq.~\eqref{eq:Ueff} we cancel off the divergent parts in $\Pi(\mathbf{q},i\nu_{n})$.
Analytically continuing the vertex function, $i\nu_{n}\rightarrow\omega+i0^{+}$,
we calculate the thermodynamic potential: 
\begin{equation}
\Omega=\Omega_{\textrm{F}}-\frac{1}{\pi}\sum_{\mathbf{q}}\int_{-\infty}^{\infty}\frac{d\omega}{e^{\beta\omega}-1}\delta(\mathbf{q},\omega),
\end{equation}
where $\delta(\mathbf{q},\omega)\equiv-{\rm Im}\ln[-\Gamma^{-1}(\mathbf{q},\omega+i0^{+})]$
is the phase of the vertex function. The number equation $nV=N=-d\Omega/d\mu$,
where $N$ is the total number of atoms and molecules and $V$ is
the area (or the volume in 2D), is solved to yield the chemical potential
$\mu$ at a given set of reduced temperature $T/T_{{\rm F}}$, binding
energy $\varepsilon_{B}/E_{{\rm F}}$, and effective range $k_{{\rm F}}^{2}R_{s}$.
Here, we define $k_{\textrm{F}}=(2\pi n)^{1/2}$, $E_{\textrm{F}}=k_{\textrm{F}}^{2}/(2M)$,
and $T_{\textrm{F}}=E_{\textrm{F}}/k_{B}$.

\begin{figure}
\centering{} \includegraphics[bb=0bp 0bp 228bp 330bp,width=0.45\textwidth]{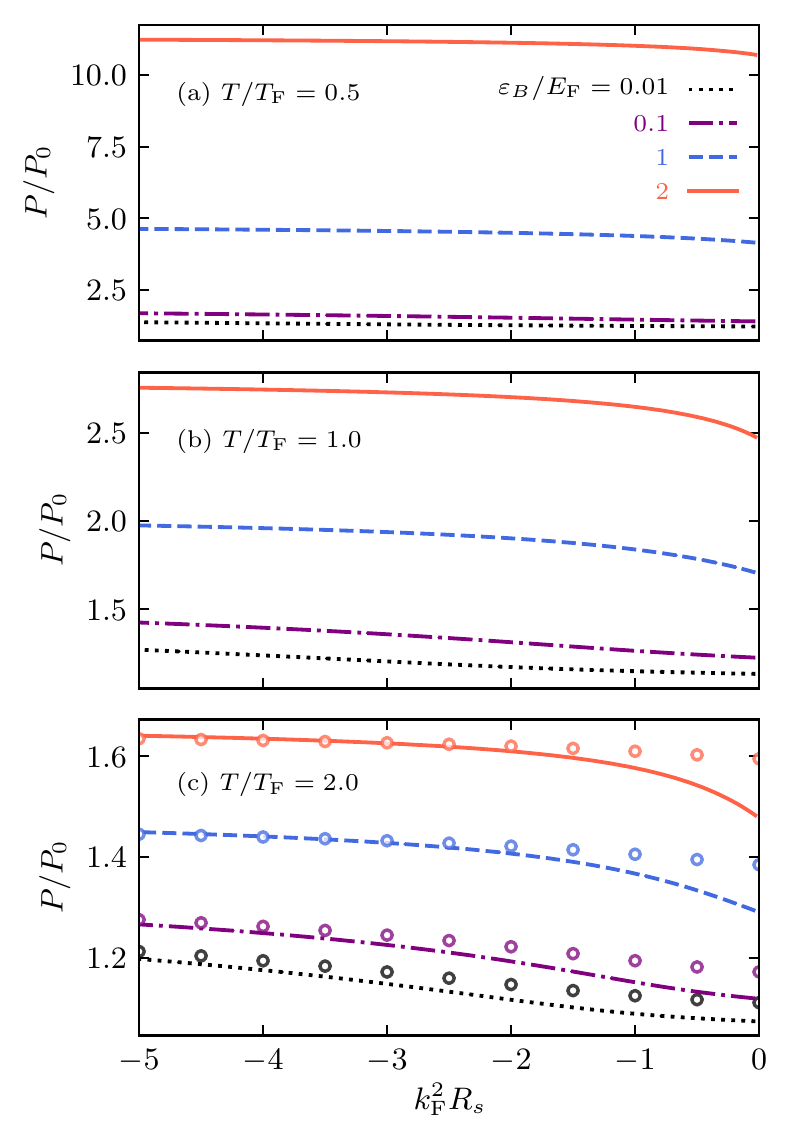}\caption{\label{fig:pressure_fixedT} Pressure EoS, normalized by the ideal
pressure $P_{0}$ at the same temperature, is plotted as a function
of the effective range for interaction strengths $\varepsilon_{B}/E_{{\rm F}}=0.01$
(black-dotted), $\varepsilon_{B}/E_{{\rm F}}=0.1$ (purple-dot dashed),
$\varepsilon_{B}/E_{{\rm F}}=1$ (blue-dashed), and $\varepsilon_{B}/E_{{\rm F}}=2$
(red-solid), at temperatures (a) $T/T_{{\rm F}}=0.5$, (b) $T/T_{{\rm F}}=1$,
and (c) $T/T_{{\rm F}}=2$. At high temperature in (c), the second
order virial expansion results for each interaction strength are shown
by circles. }
\end{figure}

\begin{figure*}
\centering{} \includegraphics[width=0.95\textwidth]{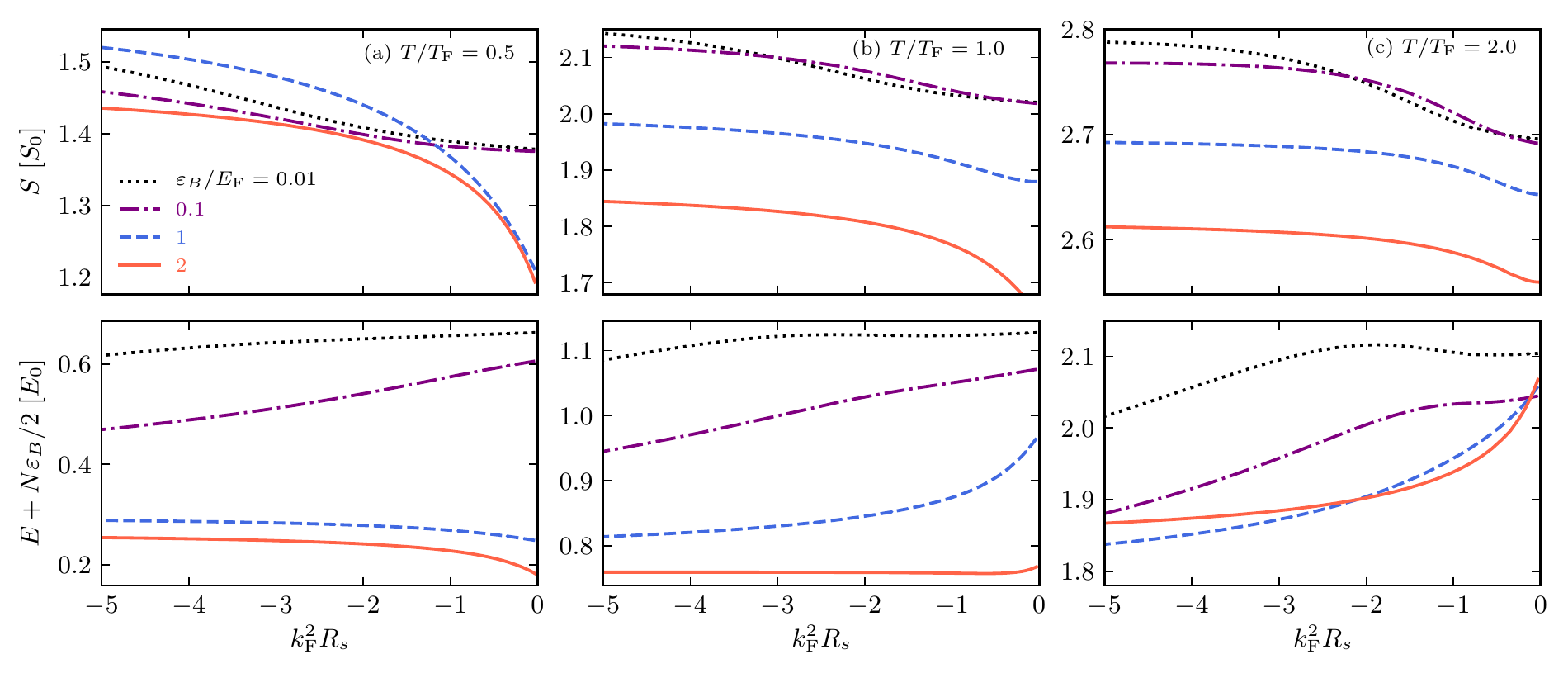}
\caption{\label{fig:thermo_fixedT} Upper panels: Entropy normalized by $S_{0}=k_{B}N$
as a function of the effective range at different interaction strengths
$\varepsilon_{B}/E_{{\rm F}}=0.01$ (black-dotted), $\varepsilon_{B}/E_{{\rm F}}=0.1$
(purple-dot dashed), $\varepsilon_{B}/E_{{\rm F}}=1$ (blue-dashed),
and $\varepsilon_{B}/E_{{\rm F}}=2$ (red-solid). The columns (a),
(b), and (c) are for temperatures $T/T_{{\rm F}}=0.5$, $T/T_{{\rm F}}=1$,
and $T/T_{{\rm F}}=2$, respectively. Lower panels: The effective
range dependence of the energy, in units of $E_{0}=NE_{{\rm F}}$,
with the same convention as in the upper panels.}
\end{figure*}

We note that within the many-body $T$-matrix framework, we cannot
calculate the superfluid transition temperature, i.e. the BKT transition
temperature. This is due to the fact that the Thouless criterion in
two dimensions becomes inapplicable as a direct consequence of Hohenberg's
theorem and the loss of long-range order due to quantum fluctuations
\cite{Loktev2001,Hohenberg1967}. The self-consistent calculation of the chemical potential and Thouless criterion always leads to a \emph{zero}
critical temperature. To consider the BKT transition 
the superfluid density needs to be calculated for the two-channel model in the superfluid
phase, following the work in Ref.~\cite{Mulkerin2017}. Such a scheme is beyond
the scope of this work.

To calculate the thermodynamic properties of the entropy and energy,
we define the dimensionless pressure equation of state $f_{p}$,
\begin{alignat}{1}
\frac{P\lambda_{T}^{2}}{k_{{\rm B}}T} & \equiv f_{p}\left(x=\frac{T}{T_{{\rm F}}},y=\frac{\varepsilon_{B}}{E_{{\rm F}}},z=k_{{\rm F}}^{2}R_{s}\right),\label{eq:Peos}
\end{alignat}
where the thermal wavelength is $\lambda_{T}=\sqrt{2\pi/(mk_{B}T)}$.
All the other thermodynamic observables can then be calculated as
derivatives from the dimensionless pressure equation of state, Eq.~\eqref{eq:Peos}.
The entropy is given by 
\begin{equation}
S=-\left(\frac{\partial\Omega}{\partial T}\right)_{\mu}\equiv Nk_{B}f_{s},
\end{equation}
where the dimensionless entropy takes the form,
\begin{equation}
f_{s}=2f_{p}-\frac{\tilde{\mu}}{x}f_{p_{x}}-\frac{k_{B}y}{x}f_{p_{y}},
\end{equation}
and $f_{p_{x}}\equiv\partial f_{p}/\partial x$, and $\tilde{\mu}=\mu/E_{{\rm F}}$.
The energy is found through $E=-TS+\Omega+\mu N\equiv NE_{{\rm F}}f_{E}$,
and we obtain 
\begin{equation}
f_{E}=xf_{s}+x\frac{f_{p}}{f_{p_{x}}}+\tilde{\mu}.\label{eq:energy}
\end{equation}

\begin{figure*}[t]
\centering{}\includegraphics[width=0.95\textwidth]{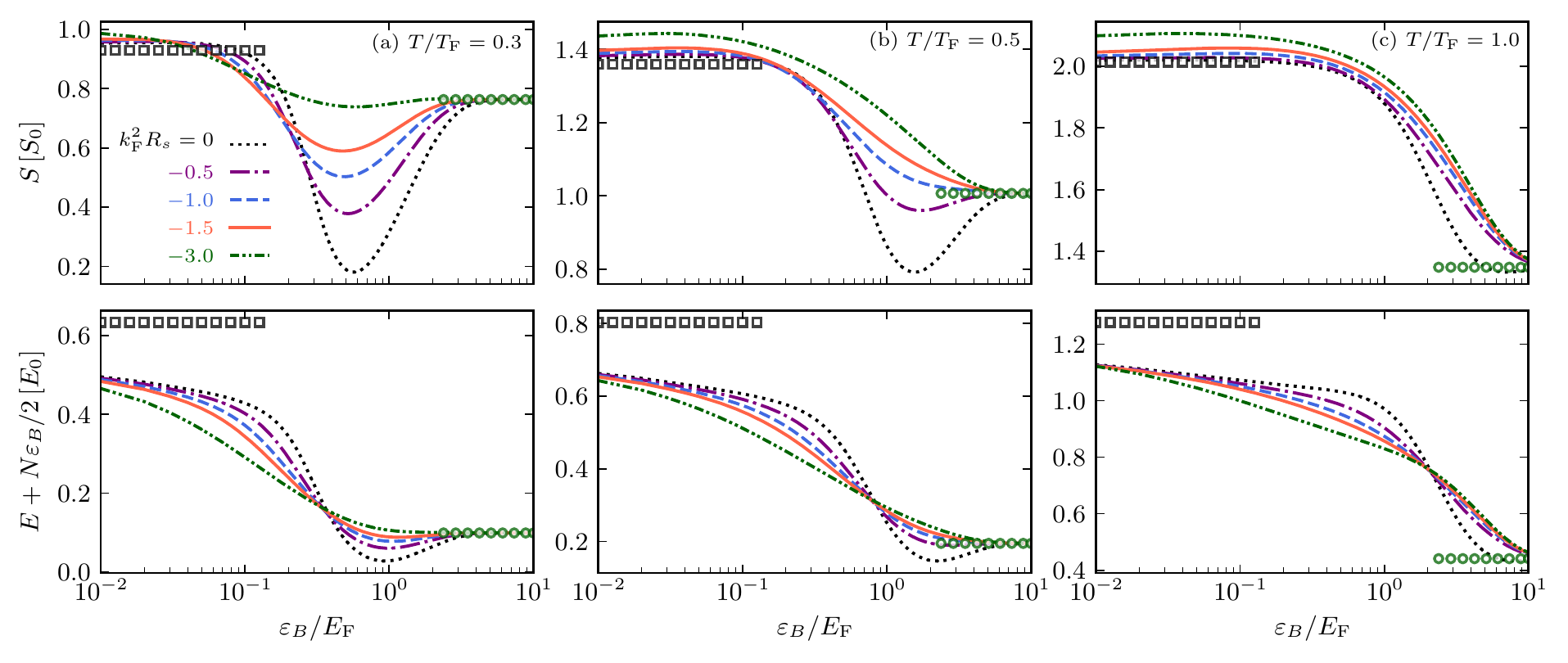}\caption{\label{fig:ES_fixedT} Upper panels: Entropy is plotted in units of
$S_{0}=Nk_{{\rm B}}$ as a function of the interaction strength at
different negative effective ranges: $k_{{\rm F}}^{2}R_{s}=0$ (black
dotted), $k_{{\rm F}}^{2}R_{s}=-0.5$ (purple dot-dashed), $k_{{\rm F}}^{2}R_{s}=-1$
(blue dashed), $k_{{\rm F}}^{2}R_{s}=-1.5$ (red solid), and $k_{{\rm F}}^{2}R_{s}=-3$
(green dot-dot-dashed), for temperatures (a) $T/T_{{\rm F}}=0.3$,
(b) $T/T_{{\rm F}}=0.5$, and (c) $T/T_{{\rm F}}=1.0$. The square
and circular symbols are the ideal Fermi and Bose gas limits (see
text), respectively. Lower panels: Energy is plotted in units of $E_{0}=NE_{{\rm F}}$.
The same convention was taken as in the upper panels.}
\end{figure*}

\section{Results and discussions}

\label{sec:results}

\subsection{Pressure equation of state}

To begin our analysis of the thermodynamic properties of the system,
we plot the pressure EoS in Fig.~\ref{fig:pressure_fixedT}, which
is normalized with the pressure of an ideal Fermi gas at the same
temperature, $P_{0}=-2\pi\lambda_{T}^{-4}{\rm Li}_{2}\left(-e^{\beta\mu}\right)$,
where ${\rm Li}_{s}(z)$ is the polylogarithm function. We show the
pressure EoS at temperatures (a) $T/T_{{\rm F}}=0.5$,
(b) $T/T_{{\rm F}}=1$, (c) and $T/T_{{\rm F}}=2$, for a range of interaction strengths,
from $\varepsilon_{B}/E_{{\rm F}}=0.01$ to $\varepsilon_{B}/E_{{\rm F}}=2$.
We find that, as the effective range decreases, the normalized pressure
EoS increases for all temperatures and interaction strengths. This
general trend is anticipated, as the system becomes less correlated
with increasing effect range. In the limit of infinitely large effective
range, $R_{s}\rightarrow\infty$ and $g\rightarrow0$, the system
simply reduces to a non-interacting mixture of atoms and molecules.

As a comparison in the high temperature regime, we consider the virial
expansion up to second order \cite{Barth2014,Liu2010,Liu2013,Ngampruetikorn2015},
\begin{equation}
f_{p}=\int_{0}^{\infty}dt\ln\left[1+ze^{-t}\right]+\Delta b_{2}z^{2}+\dots.,\label{Eq:eos}
\end{equation}
where $\Delta b_{2}$ is the second-order virial coefficient. By using
the elegant Beth-Uhlenbeck relation \cite{Beth1937}, we obtain 
\begin{equation}
\Delta b_{2}=e^{\beta\varepsilon_{B}}-\int_{0}^{\infty}\frac{dk}{k}e^{-2k^{2}}\frac{2k^{2}R_{s}+2}{\left(\ln\left[\frac{2k^{2}}{\varepsilon_{B}}\right]+k^{2}R_{s}\right)^{2}+\pi^{2}},
\end{equation}
which in the zero-range limit reduces to the known results \cite{Bauer2014,Fenech2016}. 

As shown in Fig.~\ref{fig:pressure_fixedT}(c) at a large temperature
$T=2T_{\textrm{F}}$, we find that as the effective range decreases,
the virial expansion provides an excellent description of the pressure
EoS, even for the largest binding energy. This is quite different
from the zero-range limit. At $k_{{\rm F}}^{2}R_{s}=0$, where the
single-channel model is applicable, the virial expansion systematically
overestimates the pressure EoS. The improved applicability of virial
expansion is again due to the weaker correlation at nonzero effective
range.

\subsection{Entropy and energy}

Having discussed the pressure EoS, we now consider the entropy per
particle $S/(Nk_{B})$, as shown in the upper panels of Fig.~\ref{fig:thermo_fixedT}
as a function of the negative effective range at three different temperatures,
(a) $T/T_{{\rm F}}=0.5$, (b) $T/T_{{\rm F}}=1.0$, and (c) $T/T_{{\rm F}}=2.0$,
for a range of binding energies. We find that the entropy has a consistent
behavior for all interaction strengths and temperatures: the entropy
increases as the effective range decreases. In particular, at low
temperature as the negative effective range decreases, we see for
large binding energies (i.e., $\varepsilon_{B}/E_{{\rm F}}=1$, $2$) the entropy increases rapidly. This shows that the low-temperature
entropy is more sensitive to the effective range than the pressure
EoS in the strongly interacting regime.

In the lower panels of Fig.~\ref{fig:thermo_fixedT}, we plot the
energy per particle, $E/(NE_{{\rm F}})$, with the two-body contribution
from pairs (where we have assumed that there are $N/2$ pairs each
with energy $-\varepsilon_{B}$) subtracted. For the weakest binding
energy (i.e., in the BCS regime), the energy decreases as the effective
range decreases, for all temperatures. However, for larger binding
energies at the crossover regime ($\varepsilon_{B}/E_{{\rm F}}=1,\,2$),
the energy increases at low temperature with decreasing effective
range; while at higher temperatures the energy decreases.
The different effective-range dependence of the energy at low and
high temperatures may be understood from the many-body pairing. In
the low temperature regime for large binding energy, the many-body
pairing is important and the system becomes rigid with respect to
the change of the effective range. Hence, the energy slightly increases,
similar to the pressure EoS. In contrast, at high temperature the
many-body pairing is less significant and the system experiences a
character change from atoms to molecules with increasing effective
range. The energy then decreases, roughly following the picture of
a non-interacting Bose and Fermi mixture.

To see the interaction effects on the thermodynamic properties, we
show in Fig.~\ref{fig:ES_fixedT} the dimensionless entropy (upper
panels) and energy (lower panels) as a function of the binding energy,
$\varepsilon_{B}/E_{{\rm F}}$, for a range of negative effective
ranges from $k_{{\rm F}}^{2}R_{s}=0$ to $-3$ at temperatures (a)
$T/T_{{\rm F}}=0.3$, (b) $T/T_{{\rm F}}=0.5$, and (c) $T/T_{{\rm F}}=1.0$.
For comparison, we show also the ideal gas limits of Fermi and Bose
gases for the entropy and energy. For the ideal Fermi gas limit, we
use the following formulas: 
\begin{alignat}{1}
\frac{S^{{\rm F}}}{Nk_{{\rm B}}} & =\frac{2{\rm Li}_{2}\left(-e^{\mu/E_{{\rm F}}}\right)}{{\rm Li}_{1}\left(-e^{\mu/E_{{\rm F}}}\right)}-\frac{\mu}{E_{{\rm F}}},\\
\frac{E^{{\rm F}}}{NE_{{\rm F}}} & =-\left(\frac{T}{T_{{\rm F}}}\right)^{2}{\rm Li}_{2}\left(-e^{\mu/E_{{\rm F}}}\right),
\end{alignat}
with a non-interacting chemical potential $\mu$ determined by using
the number equation. To find the entropy and energy for non-interacting
Bose gases, $S^{{\rm B}}$ and $E^{{\rm B}}$, we assume a gas of
$N/2$ non-interacting molecules with mass $2M$ and similarly solve
a molecular chemical potential.

The behavior of the entropy as a function of the binding energy is
non-trivial. We find that for $k_{{\rm F}}^{2}R_{s}=0$, as the interaction
is increased from the weakly to strongly attractive regimes, the entropy
has a local minimum at $\varepsilon_{B}/E_{{\rm F}}\simeq0.6$ for
temperature $T/T_{{\rm F}}=0.3$, $\varepsilon_{B}/E_{{\rm F}}\simeq1.05$
for temperature $T/T_{{\rm F}}=0.5$, and $\varepsilon_{B}/E_{{\rm F}}\simeq5$
for temperature $T/T_{{\rm F}}=1.0$. This minimum may be understood
as the position where pair formation is the strongest, and at low
temperature where there is a defined Fermi surface, i.e., $\mu>0$
for $\varepsilon_{B}/E_{{\rm F}}\lesssim1$, this pairing is dominated
by many-body pairing. As the negative effective range decreases, we
see this clear minimum in the crossover interaction regime becomes
shallower. For $T/T_{{\rm F}}=0.3$ the minimum shifts to weaker binding
energies with decreasing effective range, and for higher temperatures
the minimum shifts to larger binding energies and disappears for $k_{{\rm F}}^{2}R_{s}<-1$.
In the weakly attractive regime the entropy quickly approaches the
non-interacting 2D Fermi gas limit and in the strongly attractive
regime the entropy approaches the limit of $N/2$ non-interacting
molecules with mass $2M$.

The behavior of the energy is qualitatively the same as the entropy,
with a local minimum that shifts to larger binding energy as the temperature
increases. We find also that the minimum in the crossover interaction
regime becomes much weaker as the negative effective range decreases.
Unlike the entropy, the energy is slowly approaching the non-interacting
2D Fermi gas in the weakly attractive regime. However, in the strongly
attractive regime the energy quickly approaches the limit of $N/2$
non-interacting molecules.

\begin{figure}[t]
\centering{}\includegraphics[width=0.5\textwidth]{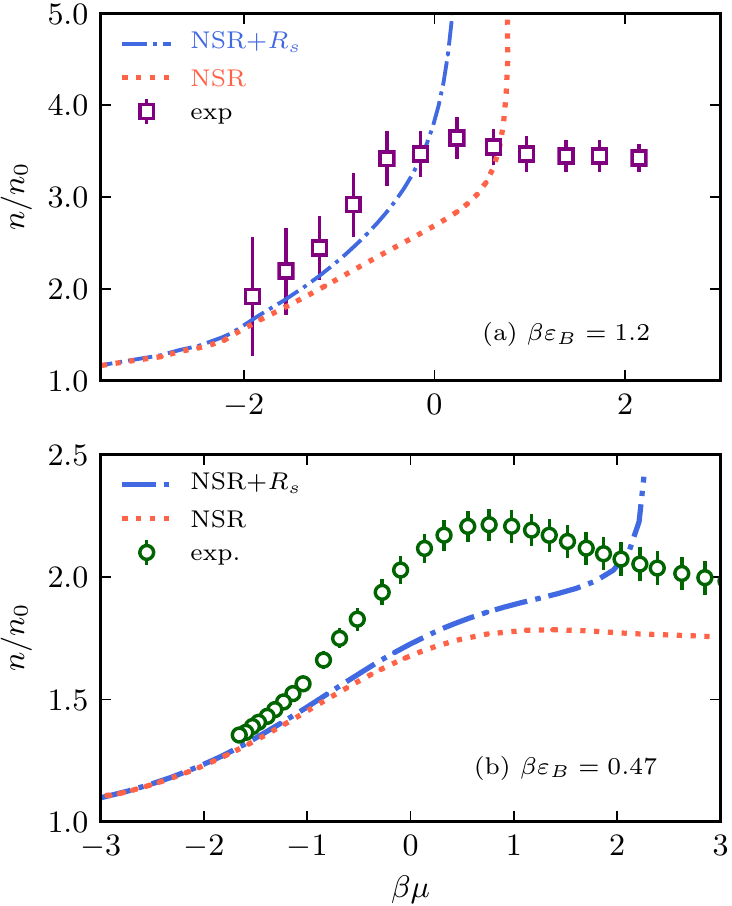}\caption{\label{fig:eos_desn120} (a) Density EoS, normalized by $n_{0}=-\lambda_{T}^{-2}{\rm Li}_{1}\left(-e^{\beta\mu}\right)$,
as a function of the dimensionless chemical potential $\beta\mu$
and interaction strength $\beta\varepsilon_{B}=1.2$, for the single-channel model (red-dotted,
$R_{s}=0$), the two-channel model (blue dot-dashed, $R_{s}\protect\neq0$),
and the experimental results from Ref.~\cite{Boettcher2016}. (b)
Density EoS at $\beta\varepsilon_{B}=0.47$, with the experimental
results from Ref.~\cite{Fenech2016}. }
\end{figure}

\subsection{Comparison to experiment}

We now compare our two-channel results to the experimental data on
density EoS, with a realistic confinement-induced effective range.
For this purpose, we match the the low-energy expansion of $T_{2B}(E^{+}\equiv k^{2}/M+i0^{+})$
to the quasi-2D scattering amplitude $f_{Q2D}(k)/M$, which describes
the two-particle scattering within the ground-state manifold under
a tight axial confinement with frequency $\omega_{z}$ \cite{Petrov2001}:
\begin{equation}
f_{Q2D}(k\rightarrow0)=\frac{4\pi}{\sqrt{2\pi}a_{z}/a_{3D}+\varpi\left(k^{2}a_{z}^{2}/2\right)},\label{eq:fQ2D}
\end{equation}
where $a_{z}\equiv\sqrt{1/(M\omega_{z})}$ is the harmonic oscillator
length and the function $\varpi(x)$ has the low-energy expansion,
$\varpi(x\rightarrow0)\simeq-\ln(2\pi x/\mathcal{B})+2x\ln2+i\pi$,
with $\mathcal{B}\simeq0.9049$.

By setting $T_{2B}(E^{+})=f_{Q2D}(k)/M$, we obtain the well-known
result \cite{Petrov2001},
\begin{equation}
a_{s}=a_{z}\frac{\pi}{\mathcal{B}}\exp\left(-\sqrt{\frac{\pi}{2}}\frac{a_{z}}{a_{3D}}\right),\label{eq:as}
\end{equation}
in the zero-energy limit $k\rightarrow0$. The determination of the
effective range $R_{s}$ is also straightforward. We require that
the two-body $T$-matrix $T_{2B}(E^{+})$ and the quasi-2D scattering
amplitude share the same pole or the same binding energy $\varepsilon_{B}$
\cite{Wu2019}. For the former, it is readily seen from Eq. (\ref{eq:T2B})
that the binding energy $\varepsilon_{B}=\kappa^{2}/M$ is related
to the effective range $R_{s}$ by,
\begin{equation}
R_{s}=\frac{2\ln\left(\kappa a_{s}\right)}{\kappa^{2}}.\label{eq:Rs}
\end{equation}
The binding energy can also be obtained from the quasi-2D scattering
amplitude by solving the equation \cite{Petrov2001}
\begin{equation}
\frac{a_{z}}{a_{3D}}=\mathcal{F}\left(\frac{\varepsilon_{B}}{\omega_{z}}\right),\label{eq:FEB}
\end{equation}
where 
\begin{equation}
\mathcal{F}(x)=\int_{0}^{\infty}\frac{du}{\sqrt{4\pi u^{3}}}\left(1-\frac{e^{-xu}}{\sqrt{\left(1-e^{-2u}\right)/(2u)}}\right).
\end{equation}
For a given $a_{z}/a_{3D}$, we can solve Eq. (\ref{eq:FEB})
for $\varepsilon_{B}=\kappa^{2}/M$, and then use Eq. (\ref{eq:Rs})
to calculate $R_{s}$. In this way, we can determine the dimensionless
ratio $R_{s}/a_{s}^{2}$ as a function of $a_{z}/a_{3D}$ \cite{Wu2019}.

Experimentally, the 2D density EoS is measured at a given magnetic field
(i.e., $a_{3D}$) and temperature through the density profile of
a cloud of $N$ interacting fermions. By applying the local density
approximation, the local density $n$ is calibrated as a function
of the normalized local chemical potential $\beta\mu$ \cite{Fenech2016,Boettcher2016}.
This gives rise to the \emph{homogeneous} density EoS $n(\mu)$ at
 dimensionless interaction parameters $\beta\varepsilon_{B}$
and $R_{s}/a_{s}^{2}$, allowing for a direct comparison with the theoretical EoS.

Figure~\ref{fig:eos_desn120}(a) plots the density EoS as a function
of dimensionless chemical potential at $\beta\varepsilon_{B}=1.2$
and $R_{s}/a_{s}^{2}\simeq-0.2$, normalized by the ideal gas density
at the same chemical potential, i.e., $n_{0}=-\lambda_{T}^{-2}{\rm Li}_{1}\left(-e^{\beta\mu}\right)$.
Here, the interaction parameters $\beta\varepsilon_{B}=1.2$ and $R_{s}/a_{s}^{2}\simeq-0.2$
are taken according to the recent density EoS measurement in Ref.~\cite{Boettcher2016}.
We plot the single-channel prediction (red-dotted line, by artificially
setting $R_{s}=0$), the two-channel prediction (blue dot-dashed line),
and the experimental data (squares). We see that including the \emph{realistic}
negative effective range quantitatively improves the comparison to
experiment in the high temperature regime, down to $\beta\mu=-0.25$
(which corresponds to a temperature $T=0.7T_{{\rm F}}$), although
as $\beta\mu$ approaches $-\infty$, the effective range has no significant
contribution due to the vanishingly small density and hence $k_{\textrm{F}}^{2}R_{s}\sim0$,
as we would expect. Typically, the normal-state NSR theory breaks
down at low temperature, as indicated by a divergent density EoS.
The NSR calculation of the two-channel model breaks down at a higher
temperature than the single-channel NSR, as a result of the fact that
the chemical potential is approaching the two-body bound state of
the closed-channel molecules.

In Fig.~\ref{fig:eos_desn120}(b) we compare the theoretical predictions
on density EoS with the experimental results of Ref.~\cite{Fenech2016}
at $\beta\varepsilon_{B}=0.47$ and $R_{s}/a_{s}^{2}\simeq-0.03$.
In this case, we are in the BCS regime with a larger 2D scattering
length $a_{s}$, so the effect of the effective range may become relatively
weaker. Although the inclusion of the effective range does significantly
shift the density EoS towards the low-temperature regime (i.e., $\beta\varepsilon_{B}\sim1$),
it does not approach the experimental results fully. This is anticipated,
as the simple many-body $T$-matrix theory such as NSR is known to
under-estimates the pair correlations close to the BKT transition.

Although our NSR theory with realistic effective ranges can not provide
a \emph{quantitative} explanation of the experimental data, the message
obtained from the comparison of the two cases (i.e., with and without
the effective range) is clear: one needs to take into account the
confinement-induced effective range in future theoretical studies
of an interacting 2D Fermi gas at finite temperature.

\section{Conclusions}

\label{sec:conc}

In summary, we have theoretically investigated the thermodynamics
of a strongly correlated Fermi gas confined to two dimensions with
a negative confinement-induced effective range. Within a two-channel
model, including pairing fluctuations beyond the mean-field level
we have discussed the negative effective range corrections to the
thermodynamic properties. Using the recent experimental data \cite{Fenech2016,Boettcher2016}
as a benchmark, we have found that density equation of state improves
in our two-channel model, compared with the widely used single-channel
model.

We have shown that as the effective range decreases, the entropy increases,
and the system eventually approaches a molecular Bose-Einstein condensate
with finite bound state energy, thus making it more difficult for
free fermions to form pairs. In future works the many-body pairing
should be carefully examined. The appearance of 2D many-body pairing
at high temperature in the crossover regime remains an interesting
and controversial topic to be explored \cite{Murthy2018,Klimin2012,Marsiglio2015,Matsumoto2018}.
\begin{acknowledgments}
Our research was supported by the Australian Research Council's (ARC)
Discovery Programs: Grant No. DP170104008 (H.H.), Grant No. FT140100003
(X.-J.L), and Grant No. DP180102018 (X.-J.L).
\end{acknowledgments}

\end{document}